\documentclass[aps,prl,floatfix,showpacs,twocolumn,superscriptaddress]{revtex4}

\usepackage{bm}
\usepackage{amssymb}
\usepackage{amsmath}
\usepackage{amstext}
\usepackage{graphics}
\usepackage{epsfig}
\usepackage{placeins}
\usepackage{psfrag}
\usepackage{subfigure}

\begin{document}

\title{Evolution from a Bose-Einstein condensate to a Tonks-Girardeau gas: \\
  An exact diagonalization study}

\author{Frank Deuretzbacher}
\email{fdeuretz@physnet.uni-hamburg.de}
\affiliation{I. Institut f\"ur Theoretische Physik, Universit\"at Hamburg,
  Jungiusstr. 9, 20355 Hamburg, Germany}

\author{Kai Bongs}
\affiliation{Institut f\"ur Laserphysik, Universit\"at Hamburg,
  Luruper Chaussee 149, 22761 Hamburg, Germany}

\author{Klaus Sengstock}
\affiliation{Institut f\"ur Laserphysik, Universit\"at Hamburg,
  Luruper Chaussee 149, 22761 Hamburg, Germany}

\author{Daniela Pfannkuche}
\affiliation{I. Institut f\"ur Theoretische Physik, Universit\"at Hamburg,
  Jungiusstr. 9, 20355 Hamburg, Germany}

\date{\today}

\begin{abstract}
  We study ground state properties of spinless, quasi one-dimensional bosons
  which are confined in a harmonic trap and interact via repulsive
  delta-potentials. We use the exact diagonalization method to analyze the
  pair correlation function, as well as the density, the momentum
  distribution, different contributions to the energy and the population of
  single-particle orbitals in the whole interaction regime. In particular, we
  are able to trace the fascinating transition from bosonic to fermi-like
  behavior in characteristic features of the momentum distribution which is
  accessible to experiments. Our calculations yield quantitative measures for
  the interaction strength limiting the mean-field regime on one side and the
  Tonks-Girardeau regime on the other side of an intermediate regime.
\end{abstract}

\pacs{03.75.Hh, 03.75.Nt, 03.65.Ge}

\maketitle

One-dimensional (1D) delta interacting bosons reveal remarkable similarities
with non-interacting fermions when the interaction between the particles is
strong \cite{Girardeau60}. The tremendous experimental progress in the field
of cold atoms has recently allowed for the realization of such strongly
interacting 1D bosonic systems \cite{Moritz03, Tolra04, Paredes04,
Kinoshita04}. The opposite regime of weak interactions is well described by
the Gross-Pitaevskii theory \cite{Gross61, Pitaevskii61}. Besides the ground
state properties, the dynamical behavior \cite{Oehberg02} in these two
limiting regimes is very different and the excitations follow the Luttinger
liquid theory \cite{Haldane81, Monien98}. Moreover, an intermediate regime is
distinguished \cite{Petrov00} by its characteristic phase fluctuations
indicating the onset of correlations between the bosons. In the homogeneous
thermodynamic limit an exact solution covering all regimes has been introduced
by E.~H.~Lieb and W.~Liniger \cite{Lieb63}, which is the basis of many
contemporary approaches \cite{Oehberg02, Dunjko01, Gangardt03, Kheruntsyan05,
Sakmann05, Hao06}. Correspondingly, most theoretical studies assume large
particle numbers. However, gases with strong correlations could only be
realized experimentally with a small number of particles so far.

In this article we concentrate on small systems with few particles where the
interaction strength between the particles can be tuned by the transverse
confinement \cite{Kinoshita04}. In particular, we study the influence of
interactions via the pair correlation function which clearly indicates the
limits of the mean-field (MF) or Bose-Einstein condensate (BEC) regime, an
intermediate regime and the Tonks-Girardeau (TG) regime. The discrimination of
these interaction regimes is an important question with relevance for current
experiments. We show that the transitions between the BEC, the intermediate
and the TG regime can also clearly be distinguished in the evolution of the
momentum distribution. The detailed analysis of the correlation function and
of the momentum distribution is a central method in the whole field of quantum
gases.

We consider spinless bosons (e.g. \textsuperscript{87}Rb atoms with frozen
spin degrees of freedom) confined in a three-dimensional cigar shaped harmonic
trap ${V_\mathrm{ext}(\vec r) = \frac{1}{2} m \omega_\bot^2 \left( x^2 + y^2
\right) + \frac{1}{2} m \omega_z^2 z^2}$; $m$ is the mass of the bosons, and
$\omega_z$ and $\omega_\bot$ are the axial and transverse angular frequencies.
The transverse confinement is much stronger than the axial confinement
${\omega_z \ll \omega_\bot}$. The bosons are assumed to interact via a delta
potential ${V_\mathrm{int}(|\vec r - \vec r\,'|) = \frac{4 \pi \hbar^2 a_s}{m}
\delta(\vec r - \vec r\,')}$, where $a_s$ is the s-wave scattering length.

The system becomes quasi one-dimensional when the transverse level spacing
${\hbar \omega_\bot}$ is much larger than the axial level spacing ${\hbar
\omega_z}$ and the three-dimensional interaction strength ${U_\mathrm{3D} =
\frac{4 \pi \hbar^2 a_s}{m l_z l_\bot^2}}$ ($l_z$, $l_\bot$: oscillator
lengths of the axial and transverse direction, $l_i = \sqrt{\frac{\hbar}{m
\omega_i}}$). Under these conditions the transverse motion in the ground state
is restricted to zero-point oscillations. Therefore, the many-particle
Hamiltonian reads
\begin{equation}
  H = \hbar \omega_z \sum_i \left( i + \frac{1}{2} \right) a_i^\dagger a_i +
  \frac{1}{2} \, U_\mathrm{1D} \sum_{ijkl} \tilde{I}_{ijkl} a_i^\dagger
  a_j^\dagger a_l a_k \; , \label{HamiltonMatrix}
\end{equation}
where $a_i^\dagger$ ($a_i$) is the bosonic creation (annihilation) operator
for a particle in an energy eigenstate $\phi_i$ of the axial harmonic
oscillator. We have neglected the constant zero mode energy of the transverse
oscillator potential. $\tilde{I}_{ijkl}$ are the dimensionless interaction
integrals ${\tilde{I}_{ijkl} = l_z \int_{-\infty}^\infty dz \phi_i(z)
\phi_j(z) \phi_k(z) \phi_l(z)}$. The essential parameter to characterize the
system is the one-dimensional interaction strength $U_\mathrm{1D}$. It results
mainly from an integration over the transverse directions. In addition, the
strong vertical confinement leads to a modification of the s-wave scattering
length \cite{Olshanii98}: ${a_\mathrm{eff} = a_s / (1 - 1.46 a_s / \sqrt{2}
l_\bot)}$. In this paper we restrict ourselves to confining frequencies
relevant to the experiments by Kinoshita {\em et al.} \cite{Kinoshita04},
${\omega_\bot = 0 \dotsc 2\pi \times 70.7 kHz}$, resulting in corrections to
$a_s$ of ${a_s < a_\mathrm{eff} < 1.16 a_s}$. These values of the transverse
frequency are far from the confinement induced resonance discussed in
Ref.~\cite{Olshanii98}. Then, ${U_\mathrm{1D} = \frac{U_\mathrm{3D}}{2 \pi}
\frac{a_\mathrm{eff}}{a_s} = 2 \sqrt{m \hbar \omega_z} a_\mathrm{eff}
\omega_\bot}$, indicating that the effective interparticle interaction can be
tuned by the transverse confinement. The Hamiltonian (\ref{HamiltonMatrix}) is
diagonalized in the subspace of the energetically lowest eigenstates of the
non-interacting many-particle system, consisting typically of up to 150000
basis states. In the following we will concentrate on results achieved for
five bosons.

\begin{figure}[t]
  \includegraphics[width = 1.0\columnwidth]{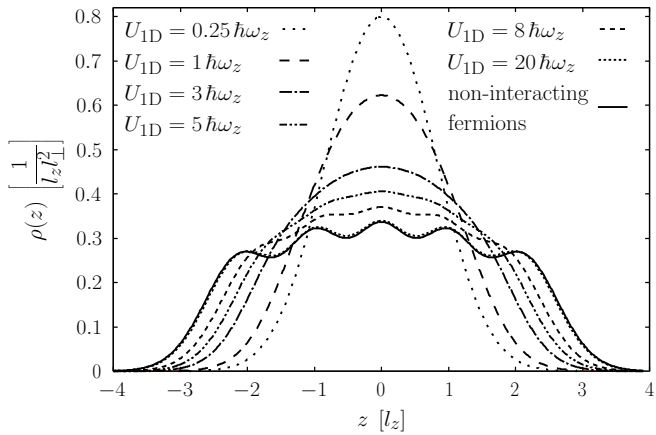}

  \caption{Particle density of five bosons for different interaction strengths
    $U_\mathrm{1D}$ along the axial direction (${x = y = 0}$). The density
    becomes flatter and broader with $U_\mathrm{1D}$. In the strong
    interaction regime it developes oscillations. At ${U_\mathrm{1D} = 20
    \hbar \omega_z}$ the density of the bosons perfectly agrees with the
    density of non-interacting fermions.
    \label{densities}}
\end{figure}

We start our discussion with the particle density ${\rho(z) = \langle \hat
\Psi^\dagger(z) \hat \Psi(z) \rangle}$ (${\hat \Psi(z)}$: field operator)
which is shown in Fig.~\ref{densities}. At small interaction strength the
density reflects the conventional mean-field behavior and ${\rho(z) \approx N
\phi_\mathrm{MF}(z)^2}$. In this regime all the bosons condense into the same
single-particle wavefunction, ${\Psi_B(z_1, ..., z_N) \approx \prod_{i = 1}^N
\phi_\mathrm{MF}(z_i)}$, and thus the many-boson system is well described by
${\phi_\mathrm{MF}(z)}$, which solves the Gross-Pitaevskii equation
\cite{Pitaevskii61, Gross61}. The system reacts to an increasing repulsive
interaction with a density which becomes broader and flatter
\cite{Kolomeisky00, Hao06, Dunjko01, Blume02}. In the strong interaction
regime density oscillations appear (see e.g. the curve at ${U_\mathrm{1D} = 8
\hbar \omega_z}$ in Fig.~\ref{densities}) and with further increasing
$U_\mathrm{1D}$ the density of the bosons converges towards the density of
five non-interacting fermions, ${\rho_F(z) = \sum_{i = 0}^4 \phi_i^2(z)}$, as
predicted by Girardeau \cite{Girardeau60}. Both densities are in perfect
agreement at ${U_\mathrm{1D} = 20 \hbar \omega_z}$ indicating that the limit
of infinite interaction is practically reached. Thus, the density oscillations
reflect the structure of the occupied orbitals in the harmonic trap. In
contrast to Ref.~\cite{Alon05} which predicts the oscillations to appear one
after the other, when the repulsion between the bosons becomes stronger, we
observe a simultaneous formation of five density maxima. These density
oscillations are absent in mean-field calculations \cite{Kolomeisky00, Dunjko01}.

\begin{figure}[t]
  \includegraphics[width = 1.0\columnwidth]{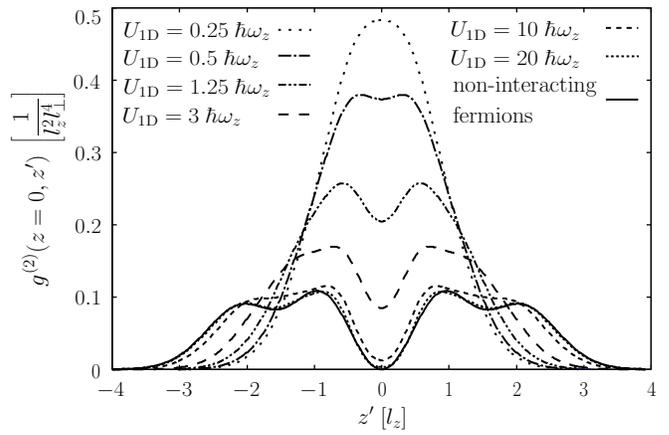}

  \caption{Pair correlation function of five bosons for different interaction
    strengths $U_\mathrm{1D}$ along the axial direction (${x = y = 0}$). One
    particle is fixed at position ${z = 0}$. The distribution flattens and
    forms a hole at coinciding particle positions, ${z = z'}$. The correlation
    function clearly indicates the transition between the three regimes (see
    text). \label{pair}}
\end{figure}

Additional insight into the evolution of the system with increasing
interaction strength can be obtained from the pair correlation function
${g^{(2)}(z, z') = \langle \hat \Psi^\dagger(z) \hat \Psi^\dagger(z') \hat
\Psi(z') \hat \Psi(z) \rangle}$ which is depicted in Fig.~\ref{pair} for
different $U_\mathrm{1D}$. In the weak interaction regime where the mean-field
approximation is valid the correlation function resembles the particle density
and ${g^{(2)}(z, z') \approx N (N - 1) \phi_\mathrm{MF}(z)^2
\phi_\mathrm{MF}(z')^2}$. The appearance of a minimum at ${U_\mathrm{1D} = 0.5
\hbar \omega_z}$ marks first deviations from this mean-field behavior. The
interparticle interaction leads to a reduced probability of finding two bosons
at the same position. These correlations are characteristic for the
intermediate regime. With increasing interaction strength the correlation hole
increases and already at ${U_\mathrm{1D} = 3 \hbar \omega_z}$ the correlation
function attains a form which is typical for a Tonks-Girardeau gas: Flat
long-range shoulders indicate the incompressibility of a Fermi gas. This
interaction strength thus  marks the transition from the intermediate to the
TG regime. By contrast, the density still exhibits a mean-field shape. The
correlation function reaches its limiting form corresponding to the one of
five non-interacting fermions at ${U_\mathrm{1D} = 20 \hbar
\omega_z}$. Besides the central correlation hole maxima appear which indicate
the position of the other particles. In our finite size system this limiting
behavior is reached at smaller interaction strength than in homogeneous systems
\cite{Astrakharchik03}. Thus, the pair correlation function clearly indicates
the limit of the mean-field regime at small interaction strength
(${U_\mathrm{1D} \leqq 0.5 \hbar \omega_z}$) and the transition from the
intermediate towards the Tonks-Girardeau regime at larger interaction strength
(${U_\mathrm{1D} \geqq 3 \hbar \omega_z}$). Within our calculations these
limits are not sensitive to the number of particles, which we have checked for
up to ${N = 7}$.

\begin{figure}[t]
  \includegraphics[width = 1.0\columnwidth]{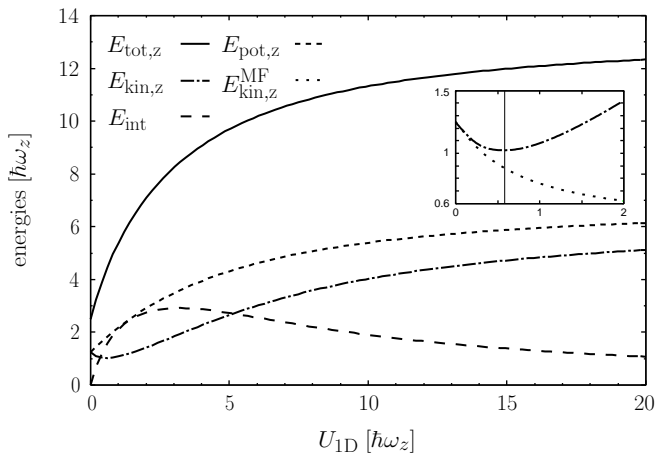}

  \caption{Evolution of various contributions to the total energy,
    $E_\mathrm{tot,z}$, of five bosons with increasing interaction strength
    $U_\mathrm{1D}$. The energies evolve towards the accordant energies of
    non-interacting fermions \cite{accuracy1}. The minimum of the kinetic
    energy coincides with the onset of significant correlations and therefore
    marks the limit of the MF regime. By contrast $E^{MF}_\mathrm{kin,z}$
    decreases in the whole interaction regime, see inset. \label{energies}}
\end{figure}

We note, that due to the singular shape of the interaction potential the
correlation function exhibits kinks at coinciding  particle positions, ${z =
z'}$, \cite{Astrakharchik03, Cirone01} which are not resolved in
Fig.~\ref{pair} \cite{accuracy2}. In recent experiments the correlation
function of three-dimensional ultracold atomic systems has been obtained by
analyzing the shot noise in absorption images \cite{Foelling05,
Greiner05}. Its value at ${z = z'}$ \cite{Gangardt03, Kheruntsyan05, Hao06}
determines, e.g., photoassociation rates \cite{Kinoshita05} and the
interaction energy. The latter is given by ${E_\mathrm{int} =
\frac{U_\mathrm{1D}}{2} \int dz g^{(2)}(z,z)}$ and is depicted in
Fig.~\ref{energies}. Its behavior is similar to the homogeneous system due to
the short range of the interaction potential \cite{Lieb63}. Nevertheless,
measurements of the whole pair correlation function are tedious. A quantity
which is easier accessible to experiments is the momentum distribution. In the
following we demonstrate that the transition between the different regimes
discussed above can also be obtained from this quantity.

We first discuss the kinetic energy, ${E_\mathrm{kin,z}= \frac{\langle p_z^2
\rangle}{2 m}}$, which is proportional to the width of the momentum
distribution. The limit of the mean-field regime at ${U_\mathrm{1D} \approx
0.5 \hbar \omega_z}$ is clearly visible in the onset of a minimum of the
correlation function (Fig.~\ref{pair}). At the same time the kinetic energy
changes dramatically, see Fig.~\ref{energies}. In the weak interaction regime
it can be approximated by ${E_\mathrm{kin,z} \approx
E^\mathrm{MF}_\mathrm{kin,z} = N \frac{\hbar^2}{2 m} \int dz \left[ \frac{d
\phi_\mathrm{MF}(z)}{dz} \right]^2}$. The flattening of the density therefore
results in the initial decrease of the kinetic energy. However, this effect is
in competition with the development of short range correlations in the
intermediate interaction regime. Strong variations of the wavefunction at
small interparticle distances lead to an increase of the kinetic energy which
can be read from the exact expression ${E_\mathrm{kin,z} = N \frac{\hbar^2}{2
m} \int dz_1 \dotsc dz_N \left[ \frac{\partial}{\partial z_1} \Psi_B(z_1
\dotsc z_N) \right]^2}$. By contrast, the mean-field kinetic energy, which is
only sensitive to variations of the density, decreases in the whole
interaction regime, see inset of Fig.~\ref{energies}. Therefore, the minimum
of the exact kinetic energy clearly marks the limit of the mean-field regime
and the dominance of interparticle correlations. With increasing particle
number $N$ the minimum of the kinetic energy slightly shifts toward smaller
values of $U_\mathrm{1D}$ with a limit at ${U_\mathrm{1D} = 0.5 \hbar
\omega_z}$ at large $N$.

We mention that the {\em potential} energy (Fig.~\ref{energies}) is
proportional to the width of  the particle density, ${E_\mathrm{pot,z} =
\frac{1}{2} m \omega_z^2 \langle z^2 \rangle}$. In the strong interaction
regime both quantities reach its limiting value of the non-interacting
fermionic system. In experiments \cite{Kinoshita04} this has been used as an
indication for the Tonks-Girardeau limit.

\begin{figure}[t]
  \includegraphics[width = 1.0\columnwidth]{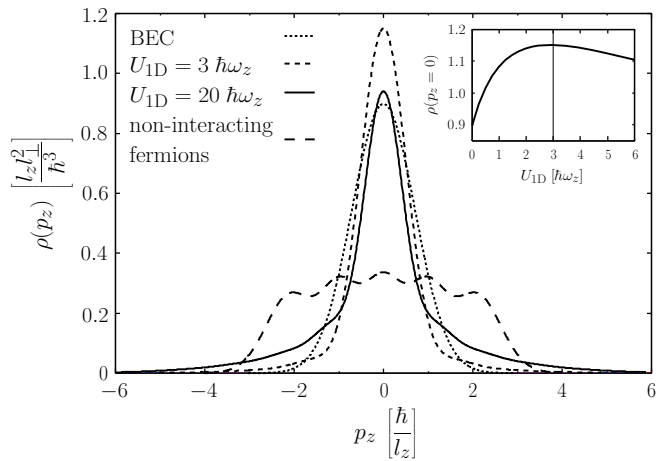}

  \caption{Momentum distribution of five bosons for different interaction
    strengths $U_\mathrm{1D}$ (${p_x = p_y = 0}$). The central peak reaches
    its maximum height at ${U_\mathrm{1D} = 3 \hbar \omega_z}$, see inset. At
    ${U_\mathrm{1D} = 20 \hbar \omega_z}$ the momentum distribution has
    developed the typical form of a harmonically trapped TG gas
    \cite{Papenbrock03} which is completely different from the distribution of
    non-interacting fermions. \label{momentum}}
\end{figure}

While the width of the momentum distribution indicates the limit of the
mean-field regime the evolution of its central peak marks the transition
towards the Tonks-Girardeau regime (Fig.~\ref{momentum}). This central result
of our studies does not depend on the number of particles. The momentum
distribution is defined as ${\rho(p_z) = \langle \hat \Pi^\dagger(p_z) \hat
\Pi(p_z) \rangle}$ where ${\hat \Pi(p_z)}$ annihilates a boson with momentum
$p_z$. Corresponding to the flattening of the density, the height of the
central maximum of the momentum distribution increases at small interaction
strength. Already in this regime, high momentum tails develop due to the kinks
in the wavefunction at coinciding particle positions, ${z_i = z_j}$
\cite{Minguzzi02, Olshanii03}. These tails are responsible for the increase of
the kinetic energy, i.e. ${\langle p_z^2 \rangle}$, at small interaction
strength above ${U_\mathrm{1D} = 0.5 \hbar \omega_z}$. However, the width of
the central peak is still shrinking in this regime, corresponding to a further
broadening of the density. By contrast, the formation of the correlation hole
at ${z = z'}$ leads to a redistribution from low towards high momentum states.
This effect dominates above ${U_\mathrm{1D} = 3 \hbar \omega_z}$, when the
growth of the density width slows down. At this point the height of the
central peak has reached its maximum. This coincides with the transition
behavior visible in the correlation function. Therefore, the interaction
strength at which the central peak of the momentum distribution reaches its
maximum height marks the transition towards the Tonks-Girardeau regime. Within
our calculations the value of ${U_\mathrm{1D} = 3 \hbar \omega_z}$ marking
this transition point is independent of the particle number $N$. The
experiments of Tolra {\it et al.} \cite{Tolra04} (${U_\mathrm{1D} \approx 4.91
\, \hbar \omega_z}$) and Kinoshita {\it et al.} \cite{Kinoshita04}
($U_\mathrm{1D}$ up to ${\approx 15.4 \, \hbar \omega_z}$), therefore, both
have been performed in the Tonks-Girardeau regime. The height of the central
peak at its maximum is approx. 30\% larger than at small interaction strength
(${U_\mathrm{1D} \approx 0}$) and about 20\% larger than at large interactions
(${U_\mathrm{1D} = 20 \hbar \omega_z}$). This contrast increases with
increasing particle number. Moreover we mention that the momentum distribution
at ${U_\mathrm{1D} = 20 \hbar \omega_z}$ perfectly agrees with exact results
obtained in the limit of infinite interaction strength \cite{Papenbrock03}.

\begin{figure}[t]
  \includegraphics[width = 1.0\columnwidth]{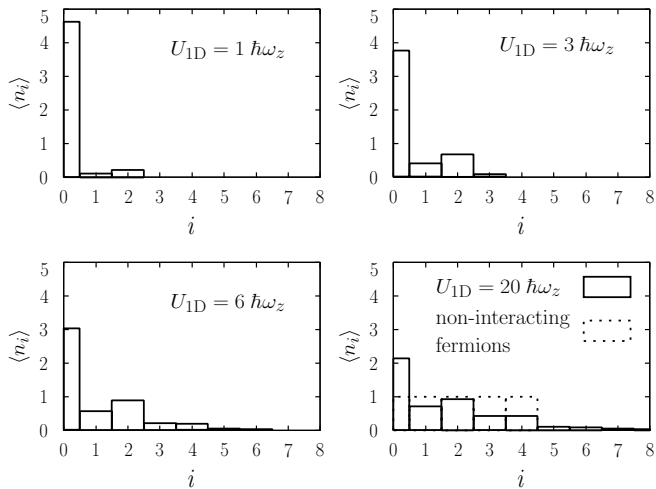}

  \caption{Occupation number distribution of five bosons for different
    interaction strengths $U_\mathrm{1D}$. With increasing $U_\mathrm{1D}$ the
    bosons leave the ground state and occupy excited states. Single-particle
    states with even parity are comparatively stronger populated than those
    with odd parity. \label{occupations}}
\end{figure}

Finally we discuss the occupation number distribution ${\langle n_i \rangle =
\langle a_i^\dagger a_i \rangle}$ of the 1D harmonic oscillator states which
is shown in Fig.~\ref{occupations}. With increasing interaction strength
$U_\mathrm{1D}$ the bosons leave the ground state and occupy excited
states. At ${U_\mathrm{1D} = 20 \hbar \omega_z}$ the distribution is similar
to the distribution shown in \cite{Girardeau01} for ${U_\mathrm{1D} =
\infty}$. However, we observe a stronger population of single-particle states
with even parity compared to those with odd parity. This effect is most
pronounced in mean-field calculations where occupations of odd parity orbitals
are absent. The comparatively stronger occupation of single-particle states
with even parity can therefore be interpreted as a remnant of the mean-field
regime.

In summary, using the exact diagonalization method we studied the
interaction-driven evolution of a quasi one-dimensional few boson
system. Besides the pair correlation function we identified the momentum
distribution as a reliable indicator for transitions of the system between
three characteristic regimes, the BEC or mean-field regime, an intermediate
regime with strong short range correlations and the Tonks-Girardeau
regime. From this we were able to quantify the interaction strength for the
transitions. The width of the momentum distribution has a minimum when the
interaction strength is approximately half as large than the axial level
spacing of the trap (${U_\mathrm{1D} = 0.5 \hbar \omega_z}$). This behavior
coincides with the onset of significant correlations and therefore marks the
transition between the mean-field and an intermediate regime. The central peak
of the momentum distribution reaches its maximum height when the interaction
strength is approximately three times larger than the axial level spacing
(${U_\mathrm{1D} = 3 \hbar \omega_z}$) and already at this point the pair
correlation function attains a form which is typical for a Tonks-Girardeau
gas. The evolution of the central peak of the momentum distribution therefore
marks the transition between the intermediate and the Tonks-Girardeau
regime. These features of the momentum distribution allow a reliable
discrimination between the three regimes. We are aware of the limitations of
our method to small particle numbers, however, we want to point out that the
method of exact diagonalization is capable to reveal the basic microscopic
mechanisms of quantum gas systems which often determine the physics of larger
systems.

\begin{acknowledgments}
  We thank S. Reimann, M. \"Ogren and H. Monien for fruitful discussions.
\end{acknowledgments}

{\it Note added in proof.} Recently, we became aware of related work by
S. Z\"ollner {\it et al.} \cite{Zoellner06_1, Zoellner06_2}.

{\bibliographystyle{apsrev}

\begin{thebibliography}{}
\bibitem{accuracy1}
The energies shown in Fig.~\ref{energies} are slightly too large since the
complete sweep has been done with a fixed basis set. One can account for the
finite basis. From this analysis we get the following true values at
${U_\mathrm{1D} = 20 \hbar \omega_z}$: ${E_\mathrm{tot} = 11.78 \hbar
\omega_z}$, ${E_\mathrm{pot} = 5.69 \hbar \omega_z}$, ${E_\mathrm{kin} = 5.07
\hbar \omega_z}$ and ${E_\mathrm{int} = 1.02 \hbar \omega_z}$. The true value
of the total energy is ${4.6 \%}$ lower than the total energy shown in
Fig.~\ref{energies}. In between, ${0 \leq U_\mathrm{1D} < 20 \hbar \omega_z}$,
the deviations are smaller.
\bibitem{accuracy2}
The exact many-particle wavefunction has kinks at coinciding particle
positions, ${z_i = z_j}$, \cite{Lieb63} which are rounded within our
approximation due to the finite size of the many-particle basis. Apart from
these deviations, which are due to the singular shape of the delta potential,
the exact wavefunction is well approximated by our solution. We checked our
method with the help of the exactly solvable two-boson problem
\cite{Cirone01}.
\end{thebibliography}

}

\end{document}